\documentclass[3p,times]{elsarticle}

\usepackage{ecrc}

\volume{00}

\firstpage{1}

\journalname{Nuclear Physics A}

\runauth{K. Hagino et al.}

\jid{NUPHA}

\jnltitlelogo{Nuclear Physics A}

\CopyrightLine{2012}{Published by Elsevier Ltd.}

\usepackage{amssymb}

\usepackage[figuresright]{rotating}
\renewcommand{\vec}[1]{\mbox{\boldmath $#1$}}

\begin{document}

\begin{frontmatter}

\dochead{}



\title{Collective excitations of $\Lambda$ hypernuclei}

\author[a]{K. Hagino}
\author[b]{J.M. Yao}
\author[c]{F. Minato}
\author[b]{Z.P. Li}
\author[a]{M. Thi Win}

\address[a]{Department of Physics, Tohoku University, Sendai 980-8578, Japan}

\address[b]{School of Physical Science and Technology, Southwest University, 
Chongqing 400715, China}

\address[c]{
Research Group for Applied Nuclear Physics, Japan Atomic Energy Agency, 
Tokai 319-1195, Japan}

\begin{abstract}
We discuss low-lying collective excitations of $\Lambda$ hypernuclei 
using the self-consistent mean-field approaches.
We first discuss 
the deformation properties of 
$\Lambda$ hypernuclei in the $sd$-shell region. 
Based on 
the relativistic mean-field (RMF) approach, 
we show that 
the oblate deformation for $^{28}$Si may disappear 
when a $\Lambda$ particle is added to this nucleus.
We then discuss the rotational
excitations of $^{25}_{~\Lambda}$Mg using the three-dimensional
potential energy surface in the deformation plane 
obtained with the Skyrme-Hartree-Fock method. 
The deformation of $^{25}_{~\Lambda}$Mg 
is predicted to be slightly reduced 
due to an addition of a $\Lambda$ particle. We demonstrate that 
this leads to a reduction of electromagnetic transition probability, $B(E2)$, 
in the ground state rotational band. 
We also present an application of random phase approximation (RPA)
to hypernuclei, and show that 
a new dipole mode, which we call a soft dipole $\Lambda$ mode, 
appears in hypernuclei, 
which can be interpreted as an oscillation
of the $\Lambda$ particle against the core nucleus.
\end{abstract}


\end{frontmatter}



\section{Introduction}

One of the main interests in hypernuclear physics is to
investigate how an addition of a $\Lambda$ particle influences the
structure of atomic nuclei. 
A characteristic feature of the $\Lambda$ particle is that it is free 
from the Pauli principle for nucleons,
and thus it can deeply penetrate into the nuclear interior.
A $\Lambda$ particle may modify several properties of nuclei, such as 
nuclear size\cite{MBI83,HK11}, the
density distribution\cite{HKYMR10},
deformation properties
\cite{Z80,ZSSWZ07,WH08,SWHS10,WHK11,YLHWZM11,LZZ11,IKDO11},
the neutron drip-line\cite{VPLR98,ZPSV08},
and fission barrier\cite{MCH09}.

In this contribution, we discuss the impurity effect of a $\Lambda$ particle 
on collective excitations. 
It is well known that low-lying states 
in even-even nuclei show a collective character, strongly reflecting the shell 
structure and pairing correlation. 
These collective excitations are one of the 
most important aspects of many-body systems, and have been extensively 
studied in the past. 
Two types of collective motion are well known: a rotational motion 
of deformed nuclei and a vibrational motion of spherical nuclei. 
In this paper, we particularly investigate the ground state rotational 
band of a deformed hypernucleus as well as the low-lying 
dipole motion of a spherical hypernucleus. 

In the next section, we first discuss the deformation properties of 
hypernuclei in the $sd$-shell region. 
In Section 3, we investigate the rotational motion of a deformed hypernucleus, 
$^{25}_{~\Lambda}$Mg. To this end, we employ the collective Hamiltonian 
approach based on the density functional theory. In Section 4, we 
discuss vibrational excitations of a spherical hypernucleus, 
$^{18}_{\Lambda\Lambda}$O. We particularly study the dipole motion, and 
show that a new dipole mode appears in hypernuclei, which is absent 
in ordinary nuclei. We then summarize the paper in Section 5. 

\section{Deformation of hypernuclei}

It is well known that many open-shell nuclei are deformed in the 
ground state. 
A clear evidence for nuclear deformation is provided by 
a rotational spectrum, which scales as $E_I\propto I(I+1)$ as a function of 
the angular momentum $I$. 
In order to investigate how the nuclear deformation is 
affected by a $\Lambda$ particle, we employ 
the self-consistent mean-field theory \cite{BHR03}. With this method, 
the optimum density distribution is obtained automatically by 
minimizing the total energy, and thus it is well suited for a discussion of 
polarization effects due to a $\Lambda$ particle. 

It is probably \v Zofka who applied the self-consistent method to deformed 
hypernuclei for the first time \cite{Z80}. 
He used Gaussian interactions for nucleon-nucleon ($NN$) and 
nucleon-Lambda ($N\Lambda$) 
interactions and showed that a $\Lambda$ particle changes 
the quadrupole moment, which is proportional to the deformation parameter, 
at most by 5 \% in the $sd$-shell region. 
This result is consistent with the results of more recent Skyrme-Hartree-Fock 
(SHF) calculations for axially deformed hypernuclei\cite{ZSSWZ07}. 

We carry out a similar study using 
the relativistic mean field (RMF) method   
as an alternative choice of effective $NN$ and $N\Lambda$ 
interactions \cite{WH08}.  
In the RMF approach, nucleons and a $\Lambda$ particle are treated as 
structureless Dirac particles interacting through the exchange of virtual 
mesons, that is, the isoscalar scalar $\sigma$ meson, 
the isoscalar vector $\omega$ meson, and the isovector vector 
$\rho$ meson\cite{VPLR98,RSM90} (see also Ref. \cite{TH12} for the zero range 
approximation to RMF, that is, the relativistic point coupling model for 
hypernuclei). 
The photon field is also taken into account to describe the Coulomb 
interaction between protons. 

\begin{figure}
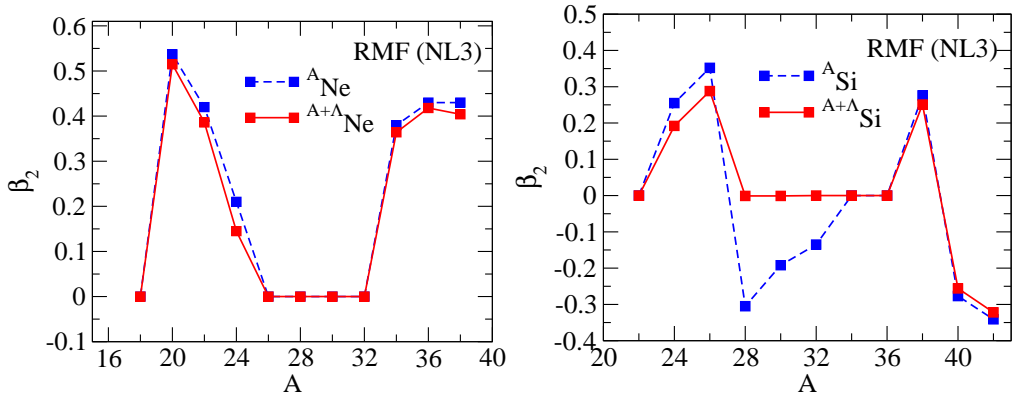

\begin{center}
\includegraphics[width=.4\textwidth,clip]{fig1a.eps}
\includegraphics[width=.4\textwidth,clip]{fig1b.eps}
\end{center}
\caption{
Quadrupole deformation parameter $\beta_2$ 
for Ne (left panel) and Si (right panel) 
isotopes obtained with the RMF method with the NL3 parameter set. 
The dashed lines show the deformation parameter for the core nucleus, while 
the solid lines for the corresponding hypernucleus. 
}
\end{figure}

Figure 1 shows the deformation parameter for the ground state 
of Ne and Si isotopes obtained with the NL3 parameter set of RMF \cite{nl3}. 
We have assumed axial symmetry for the density distribution, and 
put a $\Lambda$ particle in the lowest single-particle orbit. 
The pairing correlation among the nucleons is also  
taken into account in the constant gap approximation. 
The dashed lines show the deformation 
parameter for the even-even core nuclei, while the solid lines are for 
the corresponding hypernuclei. 
We see that 
the change in the deformation parameter 
for most of the nuclei shown in the figure is small, being consistent with 
the previous non-relativistic self-consistent 
calculations \cite{Z80,ZSSWZ07}. 
However, we find a few important exceptions. Those are 
the $^{28,30,32}$Si nuclei, for which 
the deformation parameter vanishes 
when a $\Lambda$ particle is 
added. 

\begin{figure}
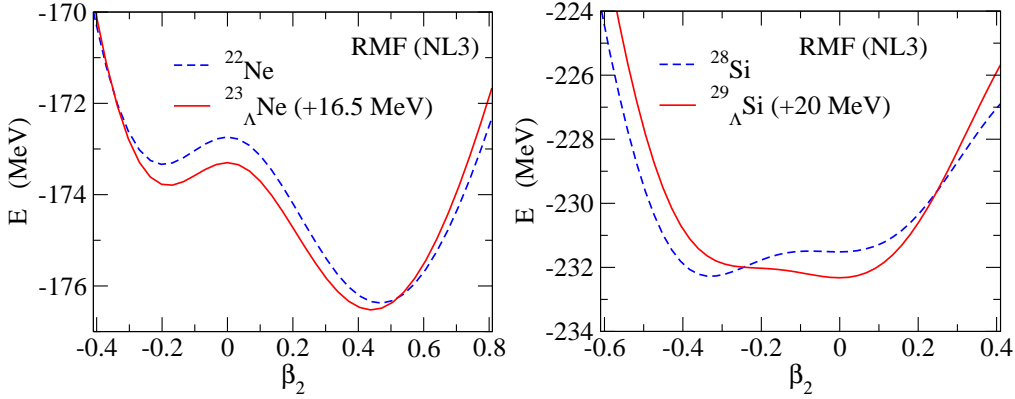

\begin{center}
\includegraphics[width=.4\textwidth,clip]{fig2a.eps}
\includegraphics[width=.4\textwidth,clip]{fig2b.eps}
\end{center}
\caption{(Left panel) 
The potential energy surface for $^{22}$Ne (the dashed line) and 
$^{23}_{~\Lambda}$Ne (the solid line) nuclei obtained with the RMF method. 
The energy surface for $^{23}_{~\Lambda}$Ne is shifted by a constant amount 
as indicated in the figure. (Right panel) The same as the left panel, but 
for $^{28}$Si and $^{29}_{~\Lambda}$Si nuclei. }
\end{figure}

In order to understand the origin for the disappearance of nuclear 
deformation, Fig. 2 shows 
the potential energy surfaces for the 
$^{23}_{~\Lambda}$Ne and $^{29}_{~\Lambda}$Si nuclei. 
The energy surfaces for the corresponding core nuclei are also shown. 
In order to facilitate the comparison, we shift the 
energy surface for the hypernuclei by a constant amount as indicated in 
the figures. 
In contrast to the $^{22}$Ne nucleus, which has a deep prolate minimum 
in the energy surface, 
the energy surface for the $^{28}$Si nucleus shows a 
relatively shallow oblate minimum, 
with a shoulder at the spherical configuration. 
The energy difference between the oblate and the spherical configurations 
is 0.754 MeV, and could be easily inverted when a $\Lambda$ particle is 
added. 

The disppearance of nuclear deformation was not observed in the 
previous SHF calculations \cite{ZSSWZ07}. 
In Ref. \cite{SWHS10}, we have compared the SHF and RMF approaches
and have shown that the different results between the two
approaches with respect to nuclear deformation come about
because the RMF yields a somewhat stronger polarization
effect of the $\Lambda$ hyperon than that of the SHF approach.
We have shown that the deformation disappears also in the
SHF approach if the energy difference between the optimum
deformation and the spherical configuration is less than about
1 MeV \cite{SWHS10}. 

Our RMF calculations indicate that the oblate 
deformation of the $^{12}$C nucleus also disappears when a $\Lambda$ particle 
is added \cite{WH08}.  It is interesting to notice that the recent 
anti-symmetrized molecular dynamics (AMD) calculation by Isaka {\it et al.} 
also exhibits a similar disappearance of deformation for the 
$^{12}$C nucleus\cite{IKDO11}. 

\section{Rotational excitations of deformed hypernuclei}

In the previous section, we presented results of RMF calculations, 
in which the axially symmetric shape of hypernuclei was assumed. 
Subsequently, three-dimensional (3D) mean-field calculations 
have also been performed both with SHF \cite{WHK11} 
and RMF \cite{LZZ11}, taking into account 
triaxial deformation, $\gamma$. 
In this section, we discuss the rotational spectra of deformed hypernuclei 
on the basis of such 3D calculations with SHF. 

In order to obtain a spectrum with the density functional theory, 
one has to go beyond the mean field approximation, in which a many-body 
wave function is assumed to be given by 
a single Slater determinant. One standard way to do so 
is to perform generator-coordinate-method (GCM) calculations, in which 
many Slater determinants are superposed after angular-momentum and 
particle-number projections \cite{BH08,YMR09,YMR10,RE10}. 
It is, however, still difficult to apply it to odd-mass nuclei, such 
as single-$\Lambda$ hypernuclei, which has a half-integer spin and  
in which the time reversal symmetry is broken. 
We therefore employ the five 
dimensional collective Bohr Hamiltonian approach \cite{NLV09,LNV09}. 
This is based on the so called Gaussian overlap approximation (GOA) to 
GCM, in which the overlap between two Slater determinants behaves as a Gaussian 
function of a generator coordinate. 

In this method, the collective Hamiltonian for a quadrupole motions, 
with the intrinsic (deformation) coordinates of $\beta$ and $\gamma$ 
together with the three-dimensional spatial rotation, is constructed as 
\begin{equation}
H_{\rm coll}=T_{\rm vib}+\frac{1}{2}\sum_{k=1}^3\frac{I_k^2}{2{\cal J}_k}
+V_{\rm coll}(\beta,\gamma),
\label{Hcoll}
\end{equation}
where $T_{\rm vib}$ is the kinetic energy operator for the vibrational motions, 
while $V_{\rm coll}$ is the collective potential. 
The vibrational moment inertia in 
$T_{\rm vib}$ and the rotational moment of inertia ${\cal J}_k$ 
in the second term of 
Eq. (\ref{Hcoll}) are calculated with the cranking approximation using 
the single-particle wave functions. 
The collective potential 
$V_{\rm coll}$ is calculated as a sum of 
the total energy in the mean-field approximation 
and corrections due to the vibrational and rotational zero point motions. 

Based on the idea of the particle-rotor/vibrator model, 
the total Hamiltonian for hypernuclei can be divided into the 
collective part for the nuclear core, the single-particle part for hyperon, 
and the interaction part between the nuclear core and hyperon\cite{BM75,RS80}. 
In this article, we focus on 
how the core nucleus is modified by the addition of a $\Lambda$ 
particle\cite{YLHWZM11}. 
We therefore apply the collective Hamiltonian approach 
only to the collective part of the Hamiltonian for the nuclear core, 
\begin{equation}
H^{(\rm core)}_{\rm coll}=
T_{\rm vib}+\frac{1}{2}\sum_{k=1}^3\frac{I_k^2}{2{\cal J}_k}
+V^{(\rm core)}_{\rm coll}(\beta,\gamma). 
\end{equation}
We mention that there is a small ambiguity here concerning how 
to define the collective potential 
for the core nucleus, $V^{(\rm core)}_{\rm coll}$, due to the $N\Lambda$ 
interaction in the energy functional. 
Here we consider two options: in 
one option, the $N\Lambda$ interaction is not included in  
$V^{(\rm core)}_{\rm coll}$ at all (this option is labeled as ``w/o'' below), 
and in the other the whole of the $N\Lambda$ interaction is included in 
$V^{(\rm core)}_{\rm coll}$ (labeled as ``w''). 

\begin{figure}
\begin{center}
\includegraphics[width=0.7\textwidth,clip]{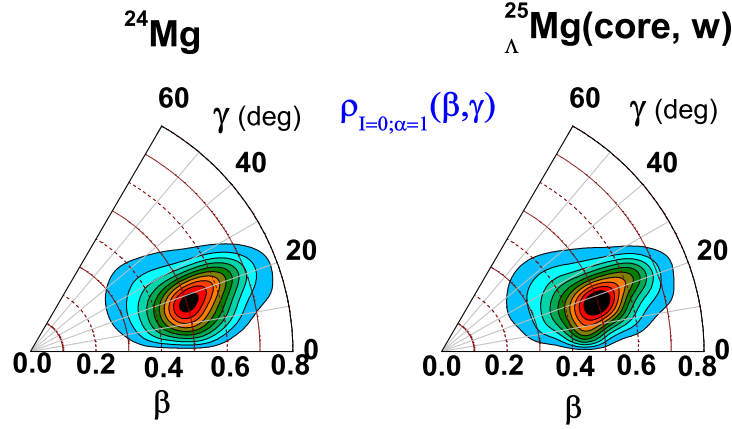}
\end{center}
\caption{
The probability distribution in the $(\beta,\gamma)$ deformation plane 
for the ground state of $^{24}$Mg without (the left panel) and with (the 
right panel) the $\Lambda$ particle. 
}
\end{figure}

By solving the collective Hamiltonian, one can construct the spectrum of 
a nucleus by taking into account the fluctuation around the minimum of 
the collective potential. 
Fig. 3 shows the probability distribution in the $(\beta,\gamma)$ 
deformation plane for the ground state of 
$^{24}$Mg when the $\Lambda$ particle is absent (the left panel) 
and present (the right panel). To this end, we use the SGII parameter 
set \cite{SG2} 
for the Skyrme interaction for the $NN$ interaction and the No. 1 set 
of Ref. \cite{YBZ88} for the $N\Lambda$ interaction. 
We take the option ``w'' for the contribution of the 
$N\Lambda$ interaction to the collective potential. 
One finds that the $\Lambda$ particle slightly shifts the 
probability distribution towards the smaller 
deformation region. 
The average values for $\beta$ and $\gamma$ are 0.54 and 20.0$^\circ$, 
respectively, for the $^{24}$Mg nucleus without $\Lambda$, which are 
altered to 0.52 and 20.8$^\circ$ when the $\Lambda$ particle is added. 
We find that the change in the proton radius is much smaller, only 
by around 0.5 \% \cite{YLHWZM11}. Therefore, the dominant effect 
of the $\Lambda$ particle in this mass region 
is to make the deformation parameter smaller, rather 
than to shrink the whole nucleus. 

\begin{figure}
\begin{center}
\includegraphics[width=0.7\textwidth,clip]{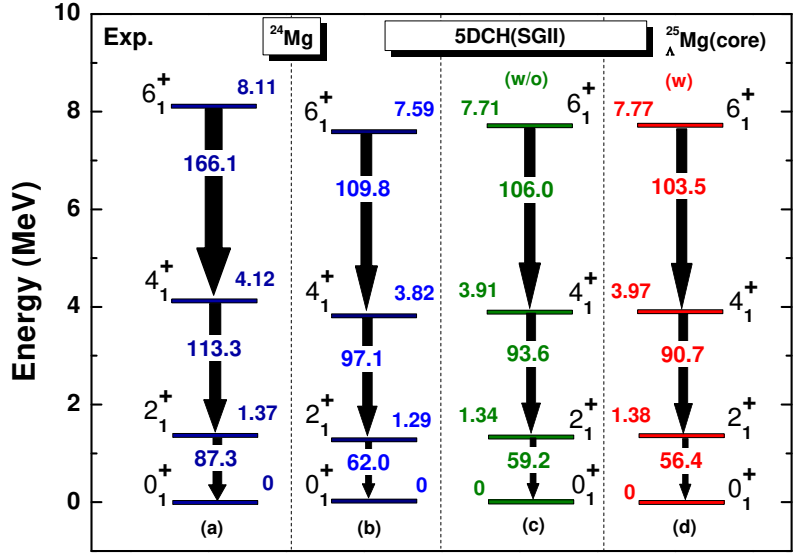}
\end{center}
\caption{
The low-spin spectra of the ground state rotational band for the
 $^{24}$Mg (b) and the nuclear core of $^{25}_\Lambda$Mg (c, d) 
obtained with the five-dimensional 
collective Hamiltonian (5DCH). 
The $B(E2)$ values are in units of e$^2$ fm$^4$.
The spectrum of $^{24}$Mg is compared with the corresponding experimental
data (a), taken from Ref.~\cite{Endt90}.
}
\end{figure}

Figure 4 shows the calculated spectra for the ground state rotational 
band. One can see that 
the $\Lambda$ stretches the spectra of ground state band. 
That is, 
the $\Lambda$ particle 
increases the excitation energy of the $2^+_1$ state by $\sim 7\%$.
At the same time, it reduces the $E2$ transition strength 
$B(E2: 2^+_1 \rightarrow 0^+_1)$ by $\sim 9\%$. 
Notice that since the $\Lambda$ particle is neutral-charge, the $E2$ 
transition strength of the whole hypernucleus is given by the protons 
in the nuclear core. Therefore, the calculated $B(E2)$ value of the 
nuclear core can be compared with the $B(E2)$ value of the hypernucleus. 
Recent AMD calculations indeed 
yield a similar reduction in the $B(E2)$ value 
for $^{25}_{~\Lambda}$Mg, although they suggest a compression 
of the rotational spectrum \cite{IHK12}. It will be an interesting future 
work to investigate how the rotational spectrum is affected when 
the $\Lambda$ particle contribution to the total Hamiltonian 
is taken into account. 

\section{Soft dipole motion of spherical hypernuclei}

Let us next discuss vibrational excitations of spherical hypernuclei. 
For vibrational excitations, 
the random phase approximation (RPA) has 
provided a convenient and useful method \cite{RS80,BB94}.
This method describes a small amplitude oscillation around the Hartree-Fock 
minimum in the potential energy surface.  
The excited phonon states are thus given in this method 
as a superposition of many 1-particle 1-hole states. 
This method has been successfully applied to many nuclei in order 
to describe low-lying collective vibrations as well as several types 
of giant resonances. See, {\it e.g.,} Ref. \cite{YN11} for a recent 
application to the giant dipole resonance in the Nd and Sm isotopes. 

One can generalize this scheme to hypernuclei \cite{MH12}. 
As in the previous section, an application to single-$\Lambda$ hypernuclei 
is complicated, and in this article we consider only double-$\Lambda$ 
hypernuclei, for which the ground state always 
has a spin and parity of 0$^+$. 
Such calculations will provide the upper limit of 
the impurity effect for single-$\Lambda$ hypernuclei. 
In RPA, excited states of hypernuclei are 
built onto the ground state $|0\rangle$ as 
$|k\rangle = Q_k^\dagger |0\rangle$ with 
\begin{equation}
Q_k^\dagger = \sum_{p,h\in n,p,\Lambda}\left(
X_{ph}^{(k)}a_p^\dagger a_h - 
Y_{ph}^{(k)}a_h^\dagger a_p\right), 
\end{equation}
where $X_{ph}^{(k)}$ and $Y_{ph}^{(k)}$ are
the forward and backward amplitudes, respectively. 
$a_p^\dagger$ and $a_h^\dagger$ are the creation operators for a 
particle state $p$ 
({\it i.e.,} a single-particle state above the Fermi energy) 
and for a hole state $h$ 
({\it i.e.,} a single-particle state below the Fermi energy), 
respectively, for protons, neutrons, and $\Lambda$ particles. 
The amplitudes $X_{ph}^{(k)}$ and $Y_{ph}^{(k)}$, as well as 
the excitation energy $E_k$ are obtained by solving 
the RPA equations, which include residual interactions. 
The electric transition probabilities from the excited state 
to the ground state are computed as 
\begin{equation}
B(E\lambda: k \to 0) = |\langle k | F |0\rangle|^2,
\end{equation}
with the excitation operators of 
\begin{equation}
F_{\lambda\mu}=e\sum_{i\in p} r_i^\lambda Y_{\lambda\mu}(\hat{\vec{r}}_i), 
\end{equation}
for $\lambda \geq 2$ and 
\begin{equation}
F_{\lambda=1,\mu}
=e\sum_{i\in p}(r_iY_{1\mu}(\hat{\vec{r}}_i)-RY_{1\mu}(\hat{\vec{R}})), 
\end{equation}
for $\lambda=1$ (that is, the E1 response), 
where 
\begin{equation}
\vec{R}=\frac{1}{m_N(Z+N)+m_\Lambda N_\Lambda}
\,\left(m_N\sum_{i\in n,p} \vec{r}_i + m_\Lambda\sum_{i\in \Lambda} 
\vec{r}_i\right), 
\end{equation}
is the center of mass of the hypernucleus,  
and 
$m_N$ and $m_\Lambda$ are 
the mass of nucleon and $\Lambda$ hyperon, respectively.
$N$, $Z$ and $N_\Lambda$ are the number of neutrons, protons and
$\Lambda$ hyperons, respectively.

\begin{table}[bt]
\begin{center}
\begin{tabular}{c|cc|cc}
\hline\hline
     & \multicolumn{2}{c|}{2$_1^+$ state} & \multicolumn{2}{c}{3$_1^-$ state}\\
\cline{2-5}
nucleus & $E$ (MeV) & $B(E2)$ (e$^2$fm$^4$) & $E$ (MeV) & $B(E3)$ (e$^2$fm$^6$)\\
\hline
$^{16}$O                   & $13.1$ & $0.726$ & $6.06$ & $91.1$ \\
$^{\,\,18}_{\Lambda\Lambda}$O & $13.8$ & $0.529$ & $6.32$ & $67.7$ \\
\hline\hline
\end{tabular}
\caption{The excitation energies and the electromagnetic transition 
probabilities, $B(E2)$ and $B(E3)$, for the first 2$^+$ and 3$^-$ states of 
$^{16}$O and $^{\,\,18}_{\Lambda\Lambda}$O nuclei obtained with the 
Skyrme-HF+RPA method.}
\end{center}
\end{table}

We apply the RPA method to the $^{18}_{\Lambda\Lambda}$O nucleus. 
To this end, 
we use the SkM$^*$ parameter set of the Skyrme interaction 
for $NN$ interaction \cite{SkM*},  
and the No.5 parameter set in Ref.\cite{YBZ88} for the $\Lambda N$
interaction. 
For the $\Lambda\Lambda$ interaction, we use the S$\Lambda\Lambda1$ parameter
set evaluated by Lanskoy\cite{Lanskoy1998}, although this parameter set may overestimate the $\Lambda\Lambda$ 
binding energy.
The results for the lowest 2$^+$ and 3$^-$ states are summarized in Table 1. 
We see that the $\Lambda$ particles slightly reduce the collectivity 
both for the quadrupole and octupole modes of excitations. That is, the 
excitation energies are increased while the electromagnetic 
transition probabilities are decreased by 26-28\%. This is qualitatively 
similar to the results for the rotational excitation presented in the 
previous section. In RPA, the increase in the excitation energies can 
be understood in terms of the change in neutron and proton single-particle 
energies due to the $\Lambda$ particles.  
That is, the $\Lambda$ particles 
affects most strongly the energy for deeply bound states, 
such as 1$s_{1/2}$ state, while 
the change is smaller for single-particle states close to the 
Fermi surface\cite{MH12}. 

\begin{figure}
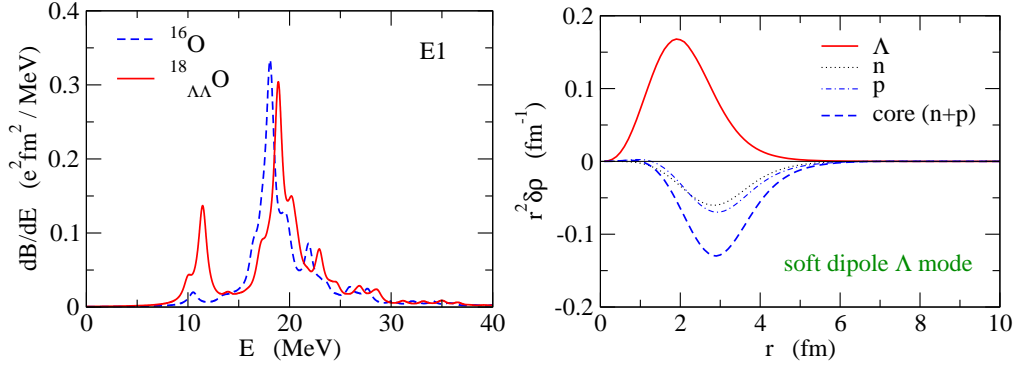

\begin{center}
\includegraphics[width=.4\textwidth,clip]{fig5a.eps}
\includegraphics[width=.4\textwidth,clip]{fig5b.eps}
\end{center}
\caption{(Left panel) 
The strength distribution for the electric dipole (E1) excitations of the 
$^{16}$O nucleus (the dashed line) and the $^{18}_{\Lambda\Lambda}$O nucleus 
(the solid line). These are obtained by smearing the RPA strength 
distributions with a Lorenzian function with a width of 1 MeV. 
(Right panel) The transition density for the soft dipole state at 
11.4 MeV in the $^{18}_{\Lambda\Lambda}$O nucleus. 
}
\end{figure}

The left panel of Fig. 5 shows 
the strength distribution for the electric dipole (E1) excitation, 
weighted by a Lorenzian function with a width of 1.0 MeV. The solid 
and dashed lines denote the results for the $^{18}_{\Lambda\Lambda}$O 
and $^{16}$O nuclei, respectively. 
This figure indicates that the addition of $\Lambda$ particles shifts 
the giant dipole resonance (GDR) peak around $\sim 18$ MeV toward a high 
energy. This feature is similar to the low-lying states shown in Table 1. 
We have found that giant resonances with other multipolarities, such as 
giant monopole resonance (GMR) and giant quadrupole resonance (GQR), 
show a similar behavior\cite{MH12}. 

In addition to the shift of the GDR peak, the dipole strength distribution 
for $^{18}_{\Lambda\Lambda}$O shows an additional peak 
at 11.4 MeV. 
This peak appears only when the $\Lambda$ hyperons are added to 
the $^{16}$O nucleus, and a similar peak is not seen in 
other modes of excitations. 
The strongest RPA amplitude, $\xi\equiv X^2-Y^2$, contributing to 
this peak is the excitation of a $\Lambda$ particle from 
the 1$s$ to the 1$p$ states with $\xi=0.873$. 
The total RPA amplitudes for the neutrons and the protons 
are small ($\xi=0.050$ for the neutrons and $\xi=0.071$ for the 
protons), and these values become entirely zero 
when the $\Lambda N$ interaction is switched off. 
The right panel of Fig. 5 
shows the transition density for this low-lying dipole state. 
The neutrons and the protons oscillate in phase, 
while they move out of phase with the $\Lambda$ particles. 
We can thus interpret this mode as a dipole oscillation of the $\Lambda$ 
particles against the core nucleus $^{16}$O. 
This is similar to the soft dipole motion 
in halo nuclei \cite{Ikeda10}, in which weakly bound valence 
neutrons oscillate against the core nucleus. 
We can therefore call the low-lying dipole peak in 
$^{18}_{\Lambda\Lambda}$O 
the soft dipole $\Lambda$ mode. 
One big difference from the soft dipole mode of halo nuclei is that 
the $\Lambda$ particles are located in the center of the hypernucleus 
whereas the valence neutrons are mainly located in the surface region in 
halo nuclei due to the Pauli principle. 
It would be an interesting future work to investigate in more details 
similarities and differences between the soft dipole mode of halo nuclei 
and that of hypernuclei. 

\section{Summary}

We have discussed the collective excitations of hypernuclei using 
self-consistent mean-field based theories. 
We have first 
investigated the 
quadrupole deformation of $\Lambda$ hypernuclei 
using the relativistic mean field (RMF) theory.
We have shown that the deformation disappears for $^{28}$Si when 
a $\Lambda$ particle is added. 
For many other nuclei, the deformation becomes slightly smaller, 
although the change is not large. 
We have demonstrated that the dominant effect of the $\Lambda$ particle 
on nuclei in the $sd$-shell region 
is to make the deformation smaller, rather than shrinking the size of 
the whole nucleus. This leads to a reduction of electromagnetic transition 
probabilities, as we have shown for the $^{25}_{~\Lambda}$Mg nucleus 
using the five dimensional collective Bohr 
Hamiltonian approach. 
We have also investigated the vibrational excitations of spherical hypernuclei 
using the random phase approximation (RPA). We have shown that a new 
low-lying dipole mode 
appears in hypernuclei, which can be interpreted as a dipole oscillation 
of $\Lambda$ particles against the core nucleus. This is a similar mode 
as the soft dipole motion in halo nuclei, and thus we call it a soft dipole 
$\Lambda$ mode. 

For both the collective Hamiltonian and RPA approaches, it is 
significantly complicated 
to apply them to single-$\Lambda$ hypernuclei. This involves 
odd-mass systems with half-integer spins and broken time reversal 
symmetry. In this article, we have avoided this difficulty by investigating 
only the core part of single-hypernuclei or by applying the theory to 
double-hypernuclei. It will be a theoretical challenge to 
develop a theory for collective excitations of single-$\Lambda$ hyerpnuclei. 
Such development will be important in view of research projects currently 
planned at the J-PARC facility using the new Ge detector array, Hyperball-J, 
that aims at obtaining new data on the low-lying energy level schemes of 
$\Lambda$ hypernuclei in the $sd$ shell region.

\section*{Acknowledgments}

We thank H. Tamura, T. Koike, H. Sagawa, and H.-J. Schulze for
useful discussions.
This work was supported by the Japanese
Ministry of Education, Culture, Sports, Science and Technology
by Grant-in-Aid for Scientific Research under
the program number (C) 22540262, 
the National Natural Science Foundation of China under Grants 
No. 11105111, No. 10947013, and No. 11105110, and 
the Fundamental Research Funds for the Central Universities 
(XDJK2010B007 and XDJK2010B007).

\end{document}